\documentclass[aip,apl,reprint]{revtex4-1}

\usepackage{graphicx}

\begin{document}

\title {Doping incorporation paths in catalyst-free Be-doped GaAs nanowires}

\author{Alberto Casadei} 
\thanks{A. Casadei and P. Krogstrup contributed equally to this work.}
\affiliation{Laboratoire des Mat\'{e}riaux Semiconducteurs, Institut des Mat\'{e}riaux, Ecole Polytechnique F\'{e}d\'{e}rale de Lausanne, CH-1015 Lausanne, Switzerland}

\author{Peter Krogstrup} 
\thanks{A. Casadei and P. Krogstrup contributed equally to this work.}
\affiliation{Nano-Science Center and Center for Quantum Devices, Niels Bohr Institute, University of Copenhagen, Denmark}

\author{Martin Heiss} 
\affiliation{Laboratoire des Mat\'{e}riaux Semiconducteurs, Institut des Mat\'{e}riaux, Ecole Polytechnique F\'{e}d\'{e}rale de Lausanne, CH-1015 Lausanne, Switzerland}

\author{Jason A. R\"{o}hr} 
\affiliation{Nano-Science Center and Center for Quantum Devices, Niels Bohr Institute, University of Copenhagen, Denmark}

\author{Carlo Colombo} 
\affiliation{Laboratoire des Mat\'{e}riaux Semiconducteurs, Institut des Mat\'{e}riaux, Ecole Polytechnique F\'{e}d\'{e}rale de Lausanne, CH-1015 Lausanne, Switzerland}

\author{Thibaud Ruelle}
\affiliation{Laboratoire des Mat\'{e}riaux Semiconducteurs, Institut des Mat\'{e}riaux, Ecole Polytechnique F\'{e}d\'{e}rale de Lausanne, CH-1015 Lausanne, Switzerland}

\author{Shivendra Upadhyay} 
\affiliation{Nano-Science Center and Center for Quantum Devices, Niels Bohr Institute, University of Copenhagen, Denmark}

\author{Claus B. S\o rensen} 
\affiliation{Nano-Science Center and Center for Quantum Devices, Niels Bohr Institute, University of Copenhagen, Denmark}

\author{Jesper Nyg\aa rd}
\affiliation{Nano-Science Center and Center for Quantum Devices, Niels Bohr Institute, University of Copenhagen, Denmark}

\author{Anna Fontcuberta i Morral}
\affiliation{Laboratoire des Mat\'{e}riaux Semiconducteurs, Institut des Mat\'{e}riaux, Ecole Polytechnique F\'{e}d\'{e}rale de Lausanne, CH-1015 Lausanne, Switzerland}

\date{\today}

\begin{abstract}
The incorporation paths of Be in GaAs nanowires grown by the Ga-assisted method in molecular beam epitaxy has been investigated by electrical measurements of nanowires with different doping profiles. We find that Be atoms incorporate preferentially via the nanowire side facets, while the incorporation path through the Ga droplet is negligible. We also demonstrate that Be can diffuse into the volume of the nanowire giving an alternative incorporation path. This work is an important step towards controlled doping of nanowires and will serve as a help for designing future devices based on nanowires.     
\end{abstract}

\maketitle

\begin{figure}[ht]
\begin{center}
\includegraphics[width=\columnwidth]{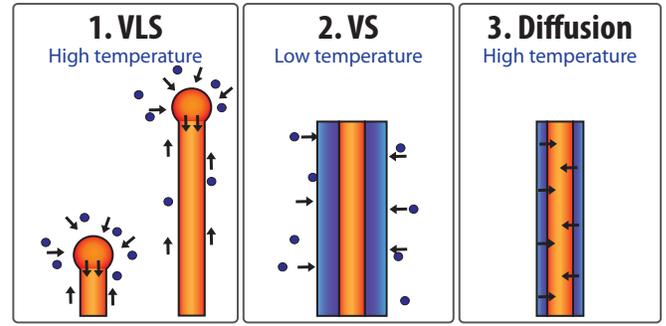}
\end{center}
\caption{Processes that influence the Be incorporation in GaAs nanowires: vapor-liquid-solid mechanism (VLS), possibility of growing a doped shell (VS) and diffusion of dopants during the growth process.}
\label{dop}
\end{figure}

Semiconductor nanowires have stimulated extensive interest in the last decade because of their potential as building blocks in future generations of electronic, optoelectronic devices as well as for energy conversion and applications \cite{The06, Cui01, Cui2001, Kel10, Cir09}. In order for nanowires to become a technological reality, control of the conductivity by doping is extremely important. Doping involves the incorporation of impurities with a small ionization energy, which can transfer carriers to either the conduction or valence band. The incorporation paths of dopants in nanowires have been discussed in the last few years, in views of controlling their concentration and position in the nanowire. For this, the understanding of the growth process of a nanowire is important. Bottom-up grown nanowires are typically formed via the Vapor-Liquid-Solid (VLS) mechanism \cite{Wag64, Mic10, Kat10}, where a nanoscale liquid droplet acts as a catalyst for nanowire crystal formation. It is generally accepted that incorporation paths of dopants are similar from the growth precursors and result either from the radial growth of the nanowire or through diffusion through the catalyst \cite{Wal11, All09, Per09, Vin09, Hil10,Ihn10}. Still, one may consider a third path, which consists on the diffusion of dopants in the core and/or from the shell to the nanowire core. The relevance of this path should depend on the coefficient of diffusion of the dopant and the growth temperatures used \cite{Yu91}. Typical dopants with a high diffusion coefficient in III-Vs are Be and Te \cite{Sal11, Cza09}. In this work, we investigate the doping of GaAs nanowires with Be. By measuring the spatial dependence of the conductivity of nanowires deposited with a flow of Be under different conditions we identify the incorporation paths and discuss the advantages and limitations of this dopant. In principle, this work can be extrapolated to other III-V materials.  \\\indent
In the correlation between the electrical conductivity and dopant incorporation, three main incorporation paths will be distinguished as schematized in Fig \ref{dop}): i) Axial incorporation through the VLS mechanism, ii) radial incorporation through the VS mechanism and as an extension of the latter and ii) diffusion of the dopants from the shell to the core in the volume of the nanowire. 

The nanowires were grown on (111) Si using a self-catalyzed (VLS)  method \cite{Col08, Ucc11, Kro10}, with a Ga deposition rate corresponding to a nominal growth of $0.27$ \AA/s , for times ranging between 30 and 60 minutes (Tab I), at 630$^\circ$C substrate temperature and a V/III flux ratio of 60\cite{Ele12}. The vertical growth rate for all the nanowires was around $15\,\mu m /h$. The p-doping was achieved by adding a flux of beryllium either during growth of the core or \textit{a posteriori} during the growth of a shell. The shell was obtained at a lower temperature (465$^\circ$C), switching the As source from As$_4$  to As$_2$  and increasing the V/III ratio to 150\cite{Col11}. To give some  insight into the Be incorporation mechanisms during axial VLS and radial VS growth of self-catalyzed GaAs nanowires, six different types of growths with varying doping profiles were carried out, see Table I. 

\begin{table}[htbp]
\centering
\begin{tabular}{lcccc}
\hline
Sample &  Growth & Nominal  & Nominal & Conductivity \\
& time & shell & conc. &  \\
& [min] & [nm] & $[atoms/cm^3]$ & $(\Omega \cdot m)^{-1}$\\
\hline
\hline
 1  & 30 & 0& $5\cdot 10^{18}$ & $0.06 \pm 0.03$\\

 2  & 60& 0 & $2.5\cdot 10^{19}$ & $16000 \pm 5000$\\

 3  & 45& 30 & $1\cdot 10^{18}$ & $3 \pm 2$\\

 4  & 45& 30 & $5\cdot 10^{18}$ & $50 \pm 20$\\

 5  & 30& 10 & $1.5\cdot 10^{19}$ & $1000 \pm 600$\\

 6  & 30& 30 & $1.5\cdot 10^{19}$ & $2200 \pm 500$\\

\hline
\end{tabular}
\caption{Time of axial  nanowires growth, nominal shell thickness, nominal doping concentration (corresponding to the planar growth under the same conditions)and measured conductivity obtained by 4-point contact configuration.}
\end{table}

\begin{figure}[ht!]
\begin{center}
\includegraphics[width=7cm]{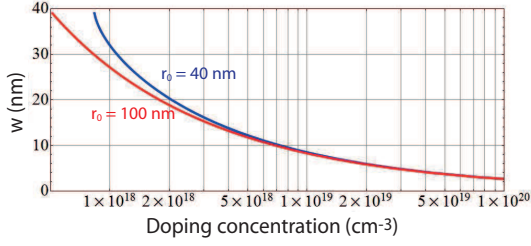}
\end{center}
\caption{Calculated depletion region width in dependence of nanowire concentration for radii $r_0=40$ nm and $r_0=100$ nm.}
\label{depl}
\end{figure}

To give a reasonable estimate of the effective carrier concentrations, the effective resistance in the nanowires was measured by carrying out 4-point electrical measurements at room temperature \cite{Duf10} on around 50 samples. The high number of devices were obtained  within a reasonable time frame with our auto-contacting software \cite{Bla12}. The electrical contacts consisted of Pd/Ti/Au (40/10/250 nm). In order to ensure ohmic and reproducible contacts, the contacted samples were annealed at 300$^\circ$C for 20 minutes.
\begin{figure}[ht]
\begin{center}
\includegraphics[width=\columnwidth]{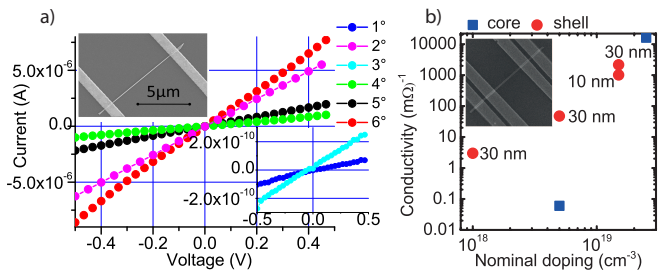}
\end{center}
\caption{a) I-V examples of each growth performed with single 2 contact configuration. The linearity of the curve shows that Pd/Ti/Au electrical contacts are ohmic on Be doped GaAs nanowires. In the insert a SEM image of a contacted nanowire. b) Nanowires conductivity obtained from 4 point measurements and a corresponding SEM image.}
\label{cond}
\end{figure}
Figure \ref{cond}a) shows a typical contacted nanowire for single 2 point contact configuration. Multiple contacts were also realized for understanding the spatial dependence of the conductivity along the nanowire axis. Examples of 2-point current-voltage characteristics are shown in the main graph of Fig. \ref{cond}a). The linearity of the curves indicates that all contacts are ohmic, even for the highly resistive samples. The conductivity of the nanowires obtained from 4-point probe measurements can be found in Tab I and in Fig. \ref{cond}b). The apparent conductivity obtained in the nanowires grown with a nominally doped core and without a shell (sample 1) is much lower as compared to the nominal doping concentration. Sample 4 has been grown with the same core conditions as Sample 1 with an additional shell of 30 nm with the same Be flux. By comparing the two samples we can separate the transport contribution due to the VLS step and obtain the contribution to the conductivity from the VS step. The doping contribution from the VLS step is negligible indicating that the flux of Be atoms to liquid-solid growth front is relatively small\cite{Gut10, Ken09}. We therefore deduce that the incorporation path of Be through the droplet can be neglected. \\\indent
The average carrier concentration in GaAs at room temperature can be obtained from the relation $p= N_{A}$, where $N_A$ is the doping density. One can use the equation:
\begin{equation}
N_A=\frac{\sigma}{\mu \cdot e}
\label{Na}
\end{equation}

where $e$ is the electrical charge of the electrons, $\mu$ is the mobility in the nanowires and $\sigma$ is the conductivity. We assume $\mu=31$ $cm^2/(V\cdot s)$, as recently measured in similar nanowires and doping concentration range \cite{Ket12}. The fundamental parameter we extract from the electrical measurements is the current crossing the nanowire. In order to extract the doping concentration, the following effects should be taken into account: (i) non-uniform radial doping distribution, (ii) surface effects such as depletion and dopant deactivation and (iii) diffusion of dopants during the growth process. Our model considers these three effects, as it is shown in the following.\\\indent
We start by presenting the effect of surface band bending, especially important for small diameters and/or low doping concentrations. The exposure of nanowires to ambient conditions causes the formation of a thin oxide layer on the surface. This results in the pinning of the Fermi level at the surface. In GaAs, the pinning occurs near the middle of the bandgap, thereby producing a depletion layer close to the surface \cite{Bjo09}. In a semi-classical model the spatial dependence of the band bending is obtained by considering the difference $\varphi$ of the Fermi level at surface states with respect to the Fermi level of the bulk. The variation of $\varphi$ in the nanowire geometry can be described by the Poisson equation in cylindrical coordinates:
\begin{equation}
\frac{1}{r}\frac{\partial}{\partial r}r\frac{\partial \varphi(r)}{\partial r}=-\frac{e N_A}{\epsilon \epsilon_0}
\end{equation}
This equation can be solved by using the boundary conditions of a vanishing electric field at the surface: $\varphi (r_0)=\Phi$, where $\Phi$ corresponds to the pinning position of the Fermi level (0.5V in p-type GaAs\cite{Lan84}).  The result is an implicit equation for the nanowire depletion layer width $w$ which depends on the nanowire radius $r_0$ and the doping concentration $N_{shell}$ \cite{Mar11}:
\begin{equation}
-\frac{w^2}{2}+r_0 w -(r_0-w)^2 ln \left( \frac{r_0}{r_0-w} \right)=\frac{2\epsilon \epsilon_0 \Phi}{e N_{shell}}
\label{depleq}
\end{equation}
The values of the depletion region $w$ as a function of the doping concentration and for two different nanowire radii is shown in Figure \ref{depl}. For high doping concentrations, the depletion region can be as small as few nm. It increases rapidly for lower doping concentrations. Already in the case of doping concentration close to $10^{18} atoms/cm^3$, the depleted region corresponds to several tens nm. For example, a nanowire with $r_0=40$ nm would be completely depleted and therefore insulating for carrier concentrations below $1\cdot 10^{18} atoms/cm^3$.

\begin{figure}[ht!]
\begin{center}
\includegraphics[width=\columnwidth]{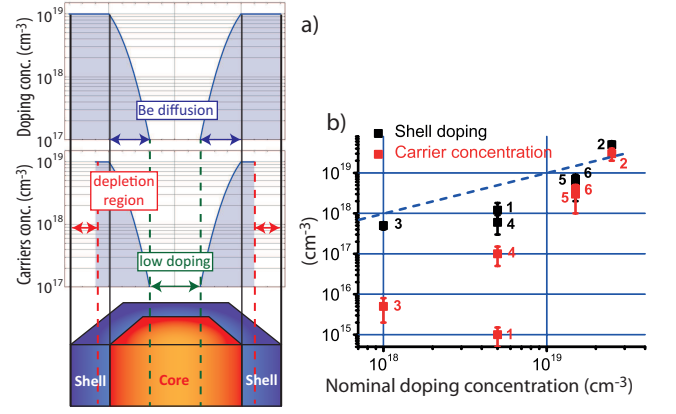}
\end{center}
\caption{a) Section view of a nanowire. The doping and the carrier concentration are reported as a function of position along the nanowire diameter. b) Comparison between carrier density concentration (red) and calculated doping concentration in the shell (black). The blue dotted line represents $N_a=N_{shell}$. The error bar reported for every dots represent only the standard deviation calculated on the electrical measurements.}
\label{d-c}
\end{figure}

The actual doping profile in the nanowire depends also on the diffusion process during growth, which is non-negligible for Be in GaAs \cite{Yu91, Ile77}. In fact, during axial growth there will be a concentration of Be adatoms on the nanowire sidewalls which can be incorporated thanks to the non-zero radial growth rate.  In fact, according to previous studies on similar nanowires \cite{Col08, Duf10}, the radial growth rate is about 1000 times lower than the axial growth rate. The concentration of Be in the vicinity of the nanowire surface is kept constant due to the constant incident Be flux, which leads to a quasi steady state concentration at the surface during growth, $p_{0}$. 
Diffusion of Be from the surface to the nanowire core is driven by the gradient in Be concentration. The diffusion is given by\cite{Kaz83}
\begin{equation}
p(x,t)=p_0\cdot erfc\left(\frac{x}{2\cdot (Dt)^{1/2}}\right)
\label{cdd}
\end{equation}
where $p_0$ is the doping concentration at the interface, $x$ is the distance from the interface, $t$ is the diffusion time and $D$ is the diffusion coefficient
\begin{equation}
D=D_0 \cdot e^{-E_0/{kT}}
\end{equation}
where for Be diffusion in GaAs, $D_0=0.655 cm^2/s$ and $E_0=2.43 eV$ \cite{Kaz83}. The diffusion length is defined as the distance with which the concentration is 1/e of the shell concentration (Table II). As schematically drawn in Fig. \ref{d-c}, the existence of a depletion region and the diffusion of Be strongly modify the range of electrically active part of the nanowires.

Since the 4-point measurements are performed in the center of the nanowire, it is assumed that the diffusion at that point corresponds to half of the full VLS growth time. The full nanowire cross-section is integrated and the depletion region included as described by equation (\ref{depleq}). Then, the doping concentration and electrical carrier concentration can be deduced from the electrical measurements:
\begin{equation}
\left\{
\begin{array}{l}
\int_0 ^{r_0} \int_0 ^{2\pi} p(x,t) x sin(\theta) dx d\theta = \frac{L}{\mu R e} \\
-\frac{w^2}{2} + r w - {r - w}^2 Log\left( \frac{r}{r - w}\right) - \frac{2 \epsilon \epsilon_0 \Phi} {e N_{shell}} = 0 
\end{array}
\right.
\end{equation}
where R/L is the resistance per unit length, $N_{shell}$ is the shell concentration and the doping distribution ($p(x,t)$) is:
\begin{equation}
p(x,t) = \left\{ \begin{array}{ll}
N_{shell}\cdot erfc\left[\frac{x}{2\cdot (Dt)^{1/2}}\right] & \mbox{core}\\
N_{shell} & \mbox{active part of the shell}\\
0 & \mbox{depletion region} 
\end{array}\right.
\end{equation}

\begin{table}[htbp]
\centering
\begin{tabular}{lccccc}
\hline
Sample  & Carrier   & Shell & Total & w &(1/e)\\
 &conc.    & conc. & shell & & Diff.\\
 &$cm^{-3}$  & $[atoms/cm^3]$& [nm]& [nm]& [nm] \\
\hline
\hline
 1   & $(1\pm 0.5)\cdot 10^{15}$  & $(1.2\pm 0.6) \cdot 10^{18}$ & 4& 27& 38 \\

 2   & $(3\pm 1)\cdot 10^{19} $  &$(5\pm 1)\cdot 10^{19} $ & 8& 3.5& 53 \\

 3   & $(5\pm 3)\cdot 10^{15} $ &$(5\pm 1)\cdot 10^{17} $ & 36& 50& 46\\

 4   & $(1\pm 0.5)\cdot 10^{17} $ &$(6\pm 3)\cdot 10^{17} $ &36& 39&46 \\

 5   & $(3\pm 2)\cdot 10^{18} $ &$(5\pm 3)\cdot 10^{18} $ & 14& 12&38\\

 6   & $(4\pm 1)\cdot 10^{18} $ &$(7\pm 2)\cdot 10^{18} $ & 34& 10&38 \\
\hline
\end{tabular}
\caption{Carrier concentration, shell doping concentration calculated with the model described previously, total expected shell, depletion region ($w$) and diffusion length defined by the distance which the concentration is 1/e of the shell concentration.}
\end{table}

For doping concentrations below $5\cdot 10^{18}$ $cm^{-3}$ the depletion region is larger than the shell thickness. This concerns sample 1,3 and 4. These three samples exhibit a very low conductivity. By taking into account the depletion region at the surface, we can extract the doping concentration at the shell and find that it is very close to the nominal doping concentration. For higher doping concentrations, the depletion region reaches few nm and affects much less the overall conductivity. We should point out that electrical measurements performed along the nanowire axis with multiple contacts showed a homogeneous conductivity along the nanowire. It is interesting to note that even though the incorporation of Be via VLS mechanism is negligible, it is possible to dope the nanowires almost homogeneously during axial growth provided the Be flux is high. For nominal doping concentration  higher than $10^{19}$ $cm^{-3}$ the highly doped shell and the long growth time results into a strong diffusion of Be inside the core, leading to an almost fully doped nanowire\cite{Kor11}. This is extremely advantageous for example in the case where an ohmic contact with a substrate is relevant (e.g. solar cells). \\\indent
In conclusion, we have shown that the Be atoms are mostly incorporated from the side facets and that the incorporation through the Ga droplet is negligible. The doping concentration is homogeneous along the nanowire and can be tuned between $6\cdot 10^{17}$ and $5\cdot  10^{19}$ $cm^{-3}$. \\\indent
The authors thank financial support from: The Swiss National Science Foundation under Grant No. 2000021-
121758/1 and 129775/1;
 NCCR QSIT; The European Research Council under Grant "Upcon";
 The Danish National Advanced
Technology Foundation through Project 022-2009-1 and the UNIK Synthetic Biology project.

\end{document}